\documentclass[letters,fleqn,usenatbib]{mnras}
\usepackage[fleqn]{amsmath}
\usepackage{graphicx}
\usepackage[varg]{txfonts}
\usepackage[T1]{fontenc}
\usepackage{ae,aecompl}
\hypersetup{pdfauthor={S. T. Ohlmann, F. K. Roepke, R. Pakmor, V. Springel, E. Mueller},pdftitle={Magnetic Field Amplification During the Common Envelope Phase}}

\newcommand{\diff}{\mathrm{d}}
\renewcommand{\vec}[1]{\boldsymbol{#1}}

\title[Magnetic Field Amplification During the CE Phase]{Magnetic Field Amplification During the Common Envelope Phase}

\author[S.~T.~Ohlmann et al.]{Sebastian~T.~Ohlmann,$^{1,2,}$%
\thanks{\texttt{sebastian.ohlmann@h-its.org}}
Friedrich~K.~R\"{o}pke,$^{1,3}$
R\"{u}diger~Pakmor,$^{1}$
\newauthor{}
Volker~Springel,$^{1,4}$
and
Ewald~M\"{u}ller$^{5}$
\\
$^{1}${%
    Heidelberger Institut f\"{u}r Theoretische Studien,
    Schloss-Wolfsbrunnenweg 35, 69118 Heidelberg, Germany
}\\
$^{2}${%
    Institut f\"{u}r Theoretische Physik und Astrophysik,
    Universit\"{a}t W\"{u}rzburg, Emil-Fischer-Str. 31, 
    97074 W\"{u}rzburg, Germany
}\\
$^{3}${%
  Zentrum f\"ur Astronomie der Universit\"at Heidelberg,
  Institut f\"ur Theoretische Astrophysik, Philosophenweg 12,
  69120 Heidelberg, Germany
}\\
$^{4}${%
  Zentrum f\"ur Astronomie der Universit\"at Heidelberg,
  Astronomisches Recheninstitut, M\"{o}nchhofstr. 12-14, 69120
  Heidelberg, Germany
}\\
$^{5}${%
  Max-Planck-Institut f\"{u}r Astrophysik, 
  Karl-Schwarzschild-Str. 1, 85748 Garching, Germany
}
}

\date{Accepted 2016 July 18. Received 2016 July 18; in original form 2016 June 14}

\pubyear{2016}

\begin{document}
\label{firstpage}
\pagerange{\pageref{firstpage}--\pageref{lastpage}}
\maketitle

\begin{abstract}
During the common envelope (CE) phase, a giant star in a binary system overflows
its Roche lobe and unstable mass transfer leads to a spiral-in of the companion,
resulting in a close binary system or in a merger of the stellar cores.
Dynamo processes during the CE phase have been proposed as a mechanism to
generate magnetic fields that are important for forming magnetic white
dwarfs (MWDs) and for shaping planetary nebulae.
Here, we present the first magnetohydrodynamics simulations of the dynamical
spiral-in during a CE phase. We find that magnetic fields are strongly amplified in the
accretion stream around the $1M_\odot$ companion as it spirals into the envelope
of a $2M_\odot$ RG\@. This leads to field strengths of 10 to 100\,kG
throughout the envelope after 120\,d. 
The magnetic field amplification is consistent with being driven by the
magnetorotational instability. The field strengths reached in our simulation
make the magnetic field interesting for diagnostic purposes, but they are
dynamically irrelevant. They are also too small
to explain the formation of the highest fields found in MWDs, but may be
relevant for luminous red novae, and detecting magnetic fields in these events
would support the scenario as proposed here.
\end{abstract}


\section{Introduction}
\label{sec:introduction}

The CE phase is usually invoked to explain the formation of
close binary systems with at least one compact star: the compact star is born
as a core of a giant with a much larger radius than the separation of the
observed binary. During the CE phase, angular momentum and energy is extracted
from the system as the envelope is ejected and a close binary system emerges.
Results of this evolution include cataclysmic variables
\citep[CVs,][]{king1988a,ritter2003a}, close WD and main sequence (MS) binaries
\citep{schreiber2003a,zorotovic2010a}, double WDs
\citep{iben1985a,han1995a,nelemans2001a}, Type Ia supernovae
\citep{iben1984a,ruiter2009a,toonen2012a}, and many more 
\citep[see also the reviews by][]{iben1993a,taam2000a,ivanova2013a}.

The origin of magnetic fields in stars is still unknown, and it is debated if
they are remainders from fields generated during the formation of a star (fossil
field hypothesis) or if they are generated during the life of a star by dynamo
processes \citep{ferrario2015a}. WDs show magnetic fields with strengths up to
$10^9$\ G and can also occur in binaries with accretion from a low-mass
companion, so-called magnetic CVs \citep{ferrario2015b}.
The absence of magnetic WDs in wide binaries suggests a binary origin for both
magnetic WDs and magnetic CVs \citep{tout2008b,briggs2015a}. In this scenario,
magnetic fields are amplified by a dynamo process in the differentially rotating
envelope during a CE phase \citep{regos1995a} or in an accretion disk around the
giant core that was formed from the tidally disrupted companion
\citep{nordhaus2011a}. The magnetic fields created during the CE phase
have to be much larger than the final surface field of the WD because it is
difficult to anchor the fields on the WD \citep{potter2010a}. At the high-mass
end of magnetic WDs, mergers of two WDs constitute a different channel that can
explain large magnetic fields \citep{zhu2015a}. In their population study,
\citet{briggs2015a} find that the major contribution comes from mergers during
the CE phase.

Magnetic fields generated during the CE phase may also affect the shape and
evolution of planetary nebulae (PNe) \citep{nordhaus2006a,nordhaus2007a}. The
magnetic field
generation relies again on a dynamo process in the differentially rotating
envelope or in an accretion disk formed by the tidally disrupted companion.
Magnetic fields have not been detected in post-AGB stars as remnants of PNe; the
upper limits lie at 100 to 300 G \citep{jordan2012c}. \citet{tocknell2014a} used
observational data on four jets in PNe to constrain the magnetic fields that are
necessary to launch these jets via the Blandford--Payne mechanism
\citep{blandford1982a} to hundreds of G to a few kG\@. Three jets
pre-date the CE event by few thousand years, one jet post-dates the CE event by
a few thousand years. Thus, besides binarity itself \citep{demarco2009a},
magnetic fields can be an important factor for shaping PNe.

Up to now, the generation of magnetic fields during the CE phase was attributed
to dynamo processes driven by shear due to differential rotation in the
envelope or in an accretion disk. Here, we present the first
magnetohydrodynamics simulations of the dynamical spiral-in during a CE phase.
The simulations extend the hydrodynamics simulation of a CE phase of a
$2M_\odot$ RG interacting with a $1M_\odot$ companion presented by
\citet{ohlmann2016a} and account for MHD effects.

\section{Methods}
\label{sec:methods}

The magnetohydrodynamics simulations presented here employ the moving-mesh code
\textsc{arepo} \citep{springel2010a} with additional modifications as introduced
in \citet{pakmor2016a} and \citet{ohlmann2016a}. As initial model, the same
$1.98M_\odot$ RG with a $0.4M_\odot$ He core is used as by \citet{ohlmann2016a}
after a relaxation step, during which spurious velocities
are damped to reach a stable equilibrium. The core of the RG and the
companion ($1M_\odot$) are represented by point masses with a gravitational
potential that is
softened at a length $h\approx 2.8R_\odot$.\footnote{This corresponds to an
  `equivalent' Plummer softening length $\epsilon \approx 1R_\odot$ (the value
  that was quoted in \citealp{ohlmann2016a}), see also the discussion of
  Eq.~(108) in \citet{springel2010a}.}
Around the point masses, an adaptive refinement is enforced with the effective
cell radius being smaller than $h/10$ in a sphere of radius
$4h$ around each point mass. During the simulation, $h$ is always enforced to be
lower than a fifth of the orbital separation. In the rest of the domain,
refinement ensures similar cell masses according to the desired resolution.  As in
\citet{ohlmann2016a}, the companion is initially placed at the surface of the
RG\@.

In addition to the simulation presented by \citep{ohlmann2016a}, magnetic fields
are included. To this end, the ideal MHD solver as implemented by
\citet{pakmor2011d} and improved by \citet{pakmor2013b} is utilized.
As initial
conditions, a magnetic dipole field was set up along the $z$ axis as
\begin{equation}
  \vec{B}(\vec{r}) = \frac{B_{\mathrm{s,i}}}{2} \frac{3\vec{n}n_z -
  \vec{e}_z}{(r/R)^3},
  \label{eq:dipole}
\end{equation}
where $B_{\mathrm{s,i}}$ denotes the initial surface field strength at the pole,
$R$ the stellar radius, $\vec{r}$ the position inside the star,
$\vec{n}=\vec{r}/r$ the normalized position, and $\vec{e}_z$ the unit vector in
$z$ direction. The initial surface field is varied in different
simulations (see Tab.~\ref{tab:simulations}); a value of $10^{-6}$\,G at the
surface of the RG would correspond to a mG surface field at the ZAMS under the
assumption of flux freezing. The initial magnetic field energy is negligible
compared to the internal energy for all models.

\section{Results}
\label{sec:results}

\subsection{Overall Evolution}
\label{sec:evolution}

\begin{figure*}
  \centering
  \includegraphics[width=\textwidth]{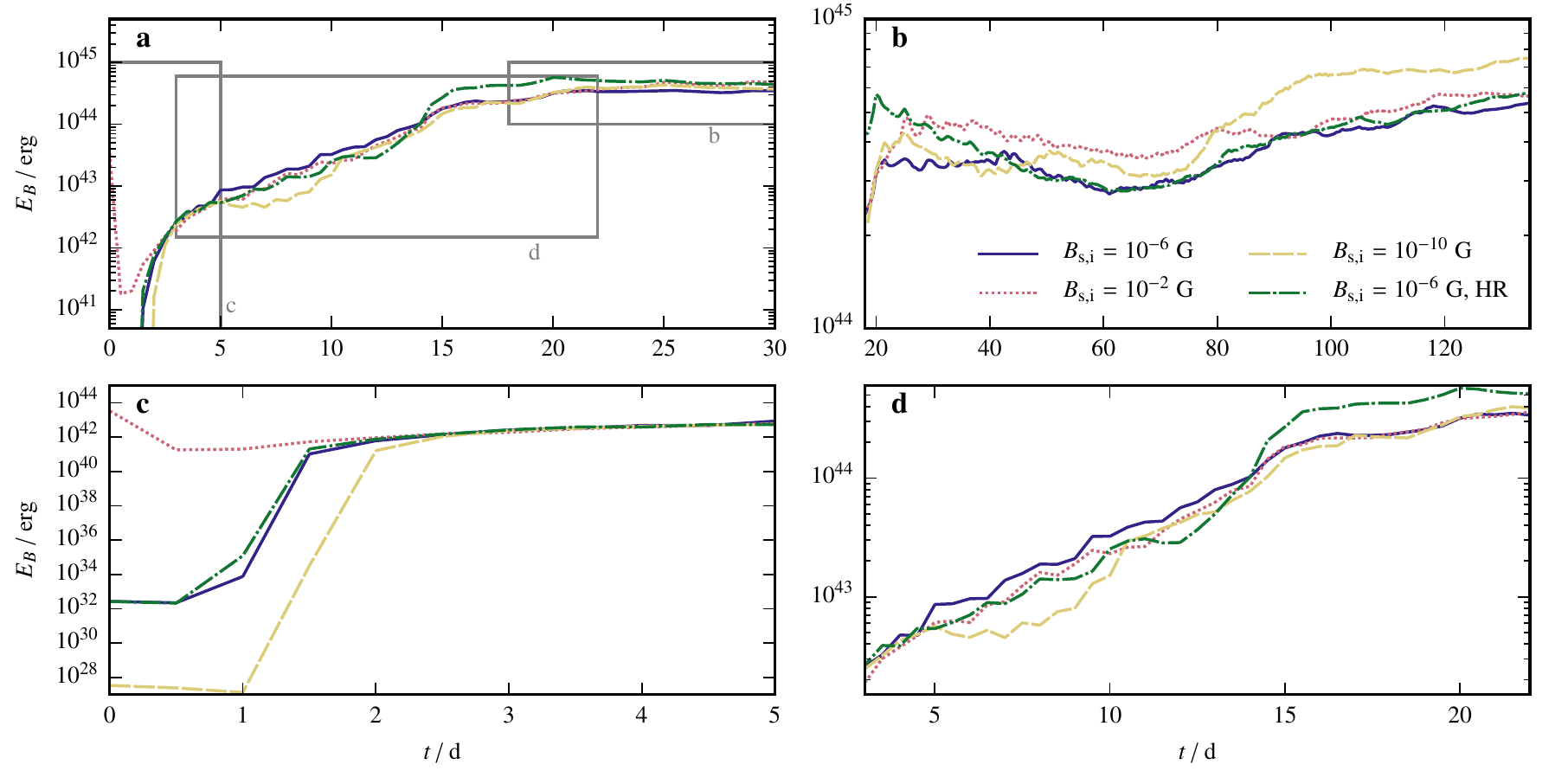}
  \vspace{-8mm}
  \caption{Magnetic field energy over time for different simulations.
    Panel~\textbf{a} displays the evolution until 30\,d,
    panel~\textbf{b} zooms in on the saturation phase during which the magnetic field
    energy stays nearly constant.  Panel~\textbf{c} shows the fast amplification
    phase during the first 5\,d of the simulations. In panel~\textbf{d}, the
    slow amplification phase is shown, which is very similar for all
    simulations. The simulations started with a different initial surface field
    $B_{\mathrm{s,i}}$. HR denotes the high-resolution simulation.}
  \label{fig:ebtime}
\end{figure*}

\begin{figure*}
  \centering
  \includegraphics[width=0.95\textwidth]{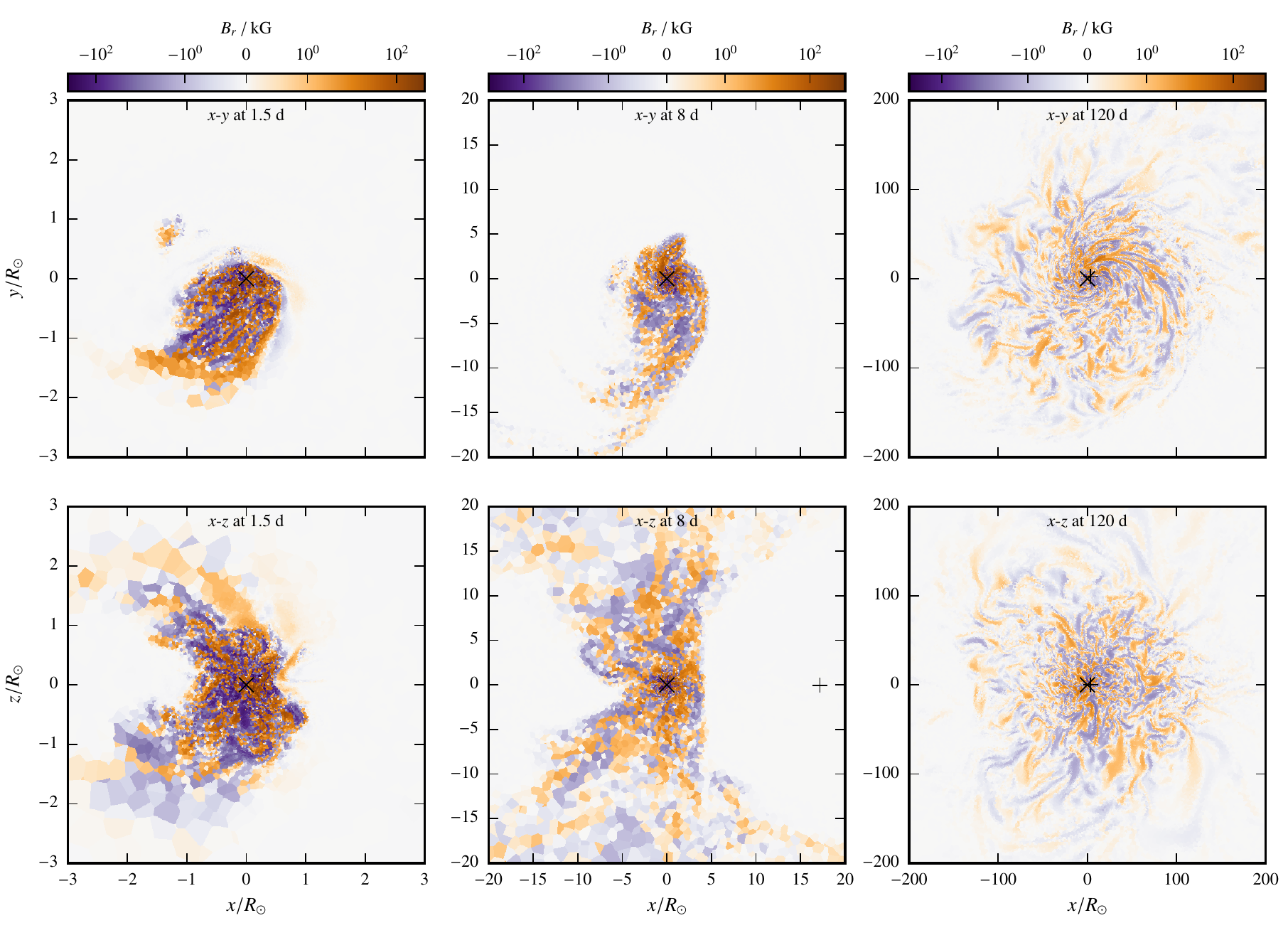}
  \vspace{-6mm}
  \caption{Magnetic field configuration over time for the high-resolution
    simulation with $B_{\mathrm{s,i}}=10^{-6}$\,G. Displayed is the radial
    magnetic field $B_r$ in the $x$-$y$ plane (upper row) and in the $x$-$z$
    plane (lower row) at three
    different times in an increasingly larger area
    centered on the companion. The position of the companion is marked by a
    $\times$, the position of the RG core is marked by a $+$; in the $x$-$z$
    plane, the position of the RG core is projected onto the plane. The RG core
    is located outside of the region shown in the two leftmost columns.  The
    color scale is linear between $-10$ and $10$\,kG and logarithmic for positive
    and negative values outside of this range.}
  \label{fig:timeseries}
  \vspace{-5mm}
\end{figure*}

The magnetic fields during the evolution are strongly amplified with a similar
evolution of the magnetic energy for all simulations
(Tab.~\ref{tab:simulations}, Fig.~\ref{fig:ebtime}) and is thus independent of
the choice of the initial surface field strength. The evolution of the magnetic
fields can be subdivided into three phases: an initial, fast amplification phase
up to 3\,d (lasting less than a fifth of the first orbit;
Fig.~\ref{fig:ebtime}, panel~c), a slow amplification phase between 3 and 20\,d
(ending after slightly more than the first orbit; Fig.~\ref{fig:ebtime},
panel~d), and a saturation phase after 20\,d (Fig.~\ref{fig:ebtime}, panel~b).
During the fast amplification phase, an accretion stream around the companion
emerges (shown in Fig.~\ref{fig:timeseries} at 1.5\,d), and the magnetic field
is amplified in this stream with an $e$-folding time of roughly 0.1\,d (cf.\
Fig.~\ref{fig:ebtime}, panel c). Regardless of the initial surface magnetic
field
($10^{-10}$, $10^{-6}$, and $10^{-2}$\,G, see Tab.~\ref{tab:simulations}), the total
magnetic energy reaches a similar value of 2-$3\times10^{42}$\,erg after this
phase at 3\,d (Fig.~\ref{fig:ebtime}, panel~c); the further evolution is very
similar for the different simulations (see panels~d and b).
For the highest surface field ($10^{-2}$\,G, red dotted line in
Fig.~\ref{fig:ebtime}), the total magnetic energy is
initially dominated by the initial field configuration in the RG, which is
unstable and decays; after 1\,d, the total magnetic energy is already dominated
by the fields around the companion (Fig.~\ref{fig:ebtime}, panel~c).  During
the slow amplification phase, the accretion stream is established and the
amplification timescale decreases to about 3\,d (see Tab.~\ref{tab:simulations};
Fig.~\ref{fig:ebtime}, panel~d; Fig.~\ref{fig:timeseries} at 8\,d). 
After 20\,d and slightly more than one orbit,
the system enters the saturation phase and the total magnetic energy remains
more or less constant at 4-$5\times10^{44}$\,erg (Fig.~\ref{fig:ebtime}, panel~b;
Fig.~\ref{fig:timeseries} at 20\,d and at 120\,d; Tab.~\ref{tab:simulations}),
which is less than a per cent of the total energy of the system
($-2.3\times10^{47}$\,erg).

As a check, the simulation with $B_{\mathrm{s,i}}=10^{-6}$\,G was repeated with
a finer mass resolution of 7.7 million cells as opposed to 2.6 million
cells for the other runs; this increase by a factor of 3 corresponds on average
to an increase in spatial resolution by a factor of about 1.4. The evolution of
this simulation is very similar to the lower-resolution simulation: the
separation does not change significantly and the total magnetic energy deviates
only between 15 and 40\,d, when it is slightly larger for the high-resolution
run (Fig.~\ref{fig:ebtime}). As demonstrated by \citet{ohlmann2016a}, it is
important to resolve the gravitational
interaction between the point masses and the gas which is ensured by requiring
an effective resolution of at least ten cells per softening length already in
the lower-resolution simulation. Increasing the resolution in the rest of the
envelope does not significantly change the outcome. Moreover, an additional
check with reduced resolution around the cores (5 cells per softening length)
does not change the outcome significantly, either. Hence, the most important
processes seem to be resolved by imposing the adaptive refinement around the
cores to at least 5 to 10 cells per softening length. 

Compared to the non-MHD run by \citet{ohlmann2016a}, the evolution of the orbit
is very similar with the final separation at 120\,d being at most 5\% larger for
the MHD simulations
(Tab.~\ref{tab:simulations}). Thus, the dynamical impact of the magnetic fields
is small, although slightly more mass is unbound during the first orbit (between
6.2\% and 6.5\% of the total envelope mass compared to 5.2\% without magnetic
fields after 60\,d), but
the difference becomes smaller at later times, when the ejected mass increases.
Moreover, the components of the magnetic stress tensor are small compared to
those of the
hydrodynamic stress tensor, and the contribution of magnetic fields to angular
momentum transport is small at all times.

The time evolution of the magnetic field structure is shown in
Fig.~\ref{fig:timeseries} in the orbital plane (upper row) and perpendicular to
it (lower row), centered on the companion. During the amplification phases, up
to 20\,d, the magnetic field strength increases in the accretion stream around
the companion and the region of very strong magnetic fields (larger than 1/10
of the maximum field strength) is confined to about $2R_\odot$ around the
companion. At the end of the slow amplification phase (Fig.~\ref{fig:timeseries}
at 20\,d), the magnetic fields start to be advected through the envelope.  At
later times (Fig.~\ref{fig:timeseries} at 120\,d), the magnetic field is
dispersed throughout the envelope, reaching values of 10 to 100\,kG in large
parts of the envelope.

\begin{table}
\centering
\caption{Simulation properties\label{tab:simulations}}
\begin{tabular}{cccccc}
\hline
{$B_{\mathrm{s,i}}$$^{a}$} &
{Cells} &
{$t_{\mathrm{amp}}$$^{b}$} & 
{$\langle E_B\rangle$$^{c}$} &
{$a_\mathrm{f}$$^{d}$} &
{$m_{\mathrm{ej}}/m_{\mathrm{env}}$$^{e}$} \\
{(G)} &
{($10^6$)} &
{(d)} & 
{($10^{44}$ erg)} &
{($R_\odot$)} &
{} \\
\hline
$10^{-6}$  & 2.6 & 3.5 & 3.9 & 4.30 & 6.5\% \\
$10^{-6}$  & 7.7 & 2.9 & 4.1 & 4.42 & 6.2\% \\
$10^{-2}$  & 2.6 & 3.3 & 4.5 & 4.33 & 6.3\% \\
$10^{-10}$ & 2.6 & 3.1 & 5.0 & 4.39 & 6.2\% \\
\hline
0$^{f}$ & 2.7 & - & - & 4.22 & 5.2\%  \\
\hline
\end{tabular}
\\
\raggedright\footnotesize
$^{a}${Initial magnetic field strength at the stellar surface.}\\
$^{b}${Amplification timescale fitted to the increase in the magnetic energy between 3 and 20\,d.}\\
$^{c}${Mean magnetic field energy between 25 and 140\,d.}\\
$^{d}${Final semi-major axis at 120\,d.}\\
$^{e}${Unbound mass over envelope mass.}\\
$^{f}${Non-MHD simulation from \cite{ohlmann2016a}.}
\end{table}

\subsection{Comparison to MRI}
\label{sec:mri}

The flow structure around the companion resembles an accretion flow with
radially decreasing angular velocity $\Omega$. In the presence of magnetic
fields, such a
configuration is unstable and subject to the magnetorotational instability
(MRI, \citealp{balbus1991a,balbus1995a}; for a review see
\citealp{balbus1998a}). During the linear phase, this instability leads to
exponentially growing magnetic fields, finally ending in a saturated state due
to non-linear interactions. The $e$-folding time during the growth phase is
usually dominated by the fastest growing MRI channel \citep[e.g.,][]{rembiasz2016a};
its growth time is given by \citep[Eq.~25]{rembiasz2016a}
\begin{equation}
  t_{\mathrm{mri}} = \frac{2}{q} \frac{1}{\Omega},
  \label{eq:tmri}
\end{equation}
and the associated wavenumber can be computed as \citep[Eq.~24]{rembiasz2016a}
\begin{equation}
  k_{\mathrm{mri}} = \sqrt{1 - \frac{2-q^2}{4}}
  \frac{\Omega}{v_{\mathrm{A}z}},
  \label{eq:kmri}
\end{equation}
where $q=-\diff \ln \Omega/\diff \ln r$ is the local rotational shear and
$v_{\mathrm{A}z} = B_z/\sqrt{4\pi\rho}$ the Alfv\'{e}n velocity in $z$ direction.
The wavelength of the corresponding MRI channel is
$\lambda_{\mathrm{mri}}=2\pi/k_{\mathrm{mri}}$.
In the simulations, $q$ takes values of about 1, thus the flow is unstable to
the MRI. The mean value of $\Omega$ around the companion rises to
$4.4\times10^{-4}$\,s$^{-1}$ during the first 0.5\,d and then slowly declines to
$5.7\times10^{-5}$\,s$^{-1}$ at 25\,d. The Alfv\'{e}n velocity increases quickly at
the beginning as the magnetic field is advected towards the companion and then
further as the MRI operates. It reaches a maximum of
$1.3\times10^{6}$\,cm\,s$^{-1}$ at 5\,d.

In the simulations, the amplification of the magnetic fields starts when the
fastest growing mode is resolved on the grid after 0.5 to 1\,d, with the
smallest grid cells having an effective radius of approximately $0.02R_\odot$.  The
$e$-folding time of the magnetic energy during the fast amplification phase
between 0.5 and 2\,d (Fig.~\ref{fig:ebtime}, panel c) is about 0.05\,d, which is
similar to the timescale of the fastest growing MRI channel (between 0.05 and
0.2\,d) as estimated as a mean value from the simulation using
Eq.~(\ref{eq:tmri}). Since the flow structure in the simulation is not an
idealized shearing box, no pure MRI channels are expected to emerge and
estimates using the formulae above are only approximate.

During the evolution, the size of the fastest growing MRI channel increases
due to an increase of the Alfv\'{e}n velocity and at 1.5\,d, it becomes larger than
the accretion flow structure, which has a radius of about
$2R_\odot$.\footnote{The magnetic field down to a tenth of the maximum field
strength is also confined to this region during the first 20\,d.} Thus, the MRI
can only operate on smaller wavelengths and longer growth times: the
$e$-folding time of the magnetic energy during the slow amplification phase is
about 3\,d (see Tab.~\ref{tab:simulations} and Fig.~\ref{fig:ebtime}, panel d),
but the fastest MRI timescale $t_{\mathrm{mri}}$ at this point is about 1\,d. It
has increased because the angular velocity  has decreased
(cf.~Eq.~\ref{eq:tmri}).

After 20\,d, the growth of the magnetic energy stops and the system transitions
to a saturation phase (Fig.~\ref{fig:ebtime}, panel b). This may be attributed
to a change in the flow structure: the accretion flow around the companion is
disturbed by the interaction with the RG core because the separation is only
$10R_\odot$ at periapsis at this time. Possible reasons usually invoked for
termination of the MRI include parasitic instabilities \citep{goodman1994a}. 
Moreover, it is
unclear what determines the level at which the magnetic fields saturate; this
may depend on the system parameters such as the evolutionary state 
(e.g., RGB or AGB)
or mass of
the primary or the initial configuration of the system. As an additional test,
we conducted a simulation with a smaller companion mass ($0.5M_\odot$), but
otherwise identical initial conditions. In this run, the same saturation level
of the magnetic energy is reached as for the $1M_\odot$ companion, thus other
factors are more important than the companion mass.

\section{Discussion}
\label{sec:discussion}

The first MHD simulations of the CE phase show that magnetic fields are
strongly amplified in the direct vicinity of the companion star on dynamical
timescales during the spiral-in.
The temporal and spatial scales are compatible with the MRI operating in the
accretion flow around the companion.  Although the magnetic field strength is
increased by orders of magnitude, the dynamical impact is small: mass loss is
slightly increased during the first orbit (about 5\% to 6\%), but the
contribution to angular momentum and energy transport is not significant and the
final separation of the stellar cores is very similar to that found in
non-MHD simulations with the same initial parameters.

This is a new way of generating magnetic fields during the dynamical spiral-in
of the CE phase compared to earlier investigations that assumed a dynamo
operating in the differentially rotating envelope \citep{regos1995a}, in an
accretion disk formed from the tidally disrupted companion
\citep{nordhaus2011a}, or in the hot outer layers of the degenerate core
\citep{wickramasinghe2014a}.
These processes, however, may still be important later
in the evolution.  We also stress that we only simulated a single system, and
that a variety of systems undergoing a CE phase has to be simulated to
understand the range of magnetic fields that can be generated during CE
evolution.

Magnetic fields that vary on short timescales, as is the case in the simulations
presented here, are difficult to anchor at the WD surface \citep{potter2010a}.
Moreover, the magnetic fields present in the simulations (10 to 100\,kG) are
smaller than the largest fields in WDs (up to $10^9$\,G, see
\citealp{ferrario2015b}). Thus, the dynamical
amplification of magnetic fields during the spiral-in may not explain the
formation of high-field magnetic WDs for the simulated CE phase of a $2M_\odot$
RG\@. According to the population synthesis calculations by \citet{briggs2015a},
the largest contribution to high-field magnetic WDs comes from mergers during
the CE phase of an AGB primary. Hence, such systems may be more promising for
future studies.

RGs can be observed with mean magnetic fields down to 10 to 100\,G, but all
observed RGs with magnetic fields cluster at the base of the RG branch
\citep{auriere2015a}. Because it is difficult to compute the photosphere in our
simulations, the magnetic field at the photosphere cannot be predicted from our
simulations, and we postpone such an analysis to future studies.
Nevertheless, CE events have been connected to luminous red novae
\citep[LRN,][]{ivanova2013b}, and the presence of magnetic fields in these
events would support dynamical amplification during the spiral-in phase of a
CE event.

\vspace{-3mm}
\section*{acknowledgments}

The work of STO was supported by Studienstiftung des deutschen Volkes
and by the graduate school GRK 1147 at University W\"urzburg.
STO and FKR acknowledge support by the DAAD/Go8 German-Australian exchange
program. RP and VS acknowledge support by the European Research Council under
ERC-StG grant EXAGAL-308037. 
STO, FKR, RP, and VS were supported by the Klaus Tschira Foundation.
EM acknowledges support from the Max-Planck-Princeton Center for Plasma
Physics.

\vspace{-3mm}
\bibliographystyle{mnras} 
\bibliography{export}

\bsp	
\label{lastpage}
\end{document}